\documentclass[journal]{IEEEtran}
\usepackage{cite}
\usepackage{graphicx}
\usepackage{amsmath} 
\usepackage{amssymb} 
\usepackage{subfigure}  
\usepackage{multirow} 
\usepackage{enumerate}
\usepackage{amsthm} 
\usepackage{tabularx} 
\usepackage{booktabs}
\usepackage{multirow}
\usepackage{array}
\usepackage{stfloats}
\usepackage{threeparttable}

\ifCLASSINFOpdf
\else
\fi

\begin{document}

\title{Power Efficiency and Delay Tradeoff of 10GBase-T Energy Efficient Ethernet Protocol}

\author{
        Xiaodan~Pan,
        Tong~Ye,~\IEEEmembership{Member,~IEEE,}
        Tony~T.~Lee,~\IEEEmembership{Fellow,~IEEE,}
        and~Weisheng~Hu,~\IEEEmembership{Member,~IEEE}

\thanks{This work was supported by the National Science Foundation of China (61571288, 61271215, and 61433009)}
\thanks{The authors are with the State Key Laboratory of Advanced Optical Communication Systems and Networks, Shanghai Jiao Tong University, Shanghai 200240, China. (email: \{pxd0506, yetong, ttlee, wshu\}@sjtu.edu.cn).}}


\maketitle

\begin{abstract}
In this paper, we study the power efficiency and delay performance of the IEEE 802.3az Energy Efficient Ethernet (EEE) protocol. A new approach is proposed to analyze the M/G/1 queue with the vacation time that is governed by the arrival process and the parameter $\tau$ and $N$ of the BTR strategy. Our key idea is to establish the connection between the vacation time and the arrival process to account for their dependency. We first derive the distribution of the number of arrivals during a vacation time based on an event tree of the BTR strategy, from which we obtain the mean vacation time and the power efficiency. Next, from the condition on the number of arrivals at the end of a vacation period, we derive a generalized P-K formula of the mean delay for EEE systems, and prove that the classical P-K formula of the vacation model is only a special case when the vacation time is independent of the arrival process. Our analysis demonstrates that the $\tau$ policy and $N$ policy of the BTR strategy are compensating each other. The $\tau$ policy ensures the frame delay is bounded when the traffic load is light, while the $N$ policy ensures the queue length at the end of vacation is bounded when the traffic load is heavy. These results, in turn, provide the rules to select appropriate $\tau$ and $N$. Our analytical results are confirmed by simulations.
\end{abstract}

\begin{IEEEkeywords}
Ethernet, IEEE802.3az, EEE protocol, M/G/1 queue with vacation, P-K formula.
\end{IEEEkeywords}

\IEEEpeerreviewmaketitle

\section{Introduction}

\IEEEPARstart{A}{s} a fundamental and pervasive network component, Ethernet has been widely applied to various kinds of networks, such as data center networks, local area networks (LANs), metropolitan area networks (MANs) and wide area networks (WANs). The number of Ethernet devices is huge \cite{Gupta:2003:GI:863955.863959}, and growing with the evolution of network technologies \cite{5522467}. Meanwhile, it is estimated that the data rate of Ethernet increases at the pace of one order of magnitude every 10 years \cite{5496875}, which significantly increases the power consumption of each Ethernet device. For example, the power consumption of a 1000Base-T Ethernet interface is only 0.5W \cite{5743052}, while that of a 10GBase-T interface is 5W \cite{kohl200710gbase}. Therefore, with the increase of the data rate, the huge number of Ethernet devices have imposed a heavy burden on the power consumption of communication networks.

On the other hand, Ethernet interfaces are idle most of the time. The link utilization of an Ethernet interface is only about 5\%-30\%, as described in \cite{1348125,nordman2007eee}. The design of the IEEE 802.3az standard \cite{5621025}, also called the Energy Efficient Ethernet (EEE) protocol, aims to relieve the power consumption of Ethernet devices. The key idea of the EEE protocol is to shut down some components of the Ethernet interface during idle periods to reduce power consumption. In other words, the interface will switch off some components, such as transceivers, and enter a so-called low power idle (LPI) mode through a Sleep operation once the transmission buffer is empty. It will wake up its components through a Wakeup operation when it accumulates enough frames in the buffer. The interface requires high power to implement the Sleep and Wakeup operations, but does not have any contributions to frame transmissions. Moreover, the durations of the Sleep and Wakeup periods are comparable to the transmission time of a frame. Thus, to maximize the power efficiency of the EEE protocol, as Fig. \ref{EEEpro} illustrates, the frequency of the Sleeps and the Wakeups should be limited.

Recently, a wakeup strategy called burst transmission (BTR) \cite{5432134,Kim2013350} has attracted a lot of interest. The BTR strategy does not trigger the Wakeup at once when the first frame arrives after the beginning of the Sleep. Instead, the BTR strategy initializes a timer and a counter upon the first arrival and triggers the Wakeup when the timer or the counter reaches a predetermined threshold. The BTR strategy ensures that the duration of the LPI mode is sufficiently long, and the number of frames that can be transmitted after each Wakeup is sufficiently large, such that the power efficiency can be higher. However, when these thresholds become too large, the BTR strategy may worsen the delay performance, since the interface is not allowed to transmit the frames that arrive during idle periods. This implies that there is a tradeoff between the power efficiency and the queueing delay. The purpose of this paper is to study the influence of the two important BTR parameters, the timer threshold $\tau$ and the counter threshold $N$, on the system performance.

\subsection{Previous work}
Several analytical models have been developed to study the EEE protocol in recent years. The simplest case is considered in \cite{Larrabeiti2011131,5871409,6689480}, in which $\tau=0$ and $N=1$, that is, the Wakeup is triggered by the first arrival frame. This kind of BTR strategy is also called frame transmission (FTR) scheme \cite{Kim2013350}. The time is slotted according to the frame transmission time and a discrete-time Markov chain is developed in Ref. \cite{Larrabeiti2011131}, but it is very difficult to solve for a closed-form solution. Thus, this paper only obtains the power efficiency and does not derive the mean delay. Ref. \cite{5871409} derives the power efficiency through a simple model, which cannot be extended to study the mean delay of a frame. To obtain both the power consumption and mean delay, a thorough analytical framework for the EEE systems with the link rate ranging from 100Mb/s to 10Gb/s is proposed in Ref. \cite{6689480}.However, previous results reported in \cite{5432134,5282379} show that the FTR strategy does not work well and $\thicksim$86\% power is consumed by the Sleep and the Wakeup operations in the worst case.

Thus, Ref. \cite{6196920} uses a deterministic model to estimate the performance of the BTR strategy with $\tau\rightarrow0$ and $N>1$, that is, only the counter is used. Despite this model obtaining both the power efficiency and the mean waiting time, it is less accurate when $N$ is small. On the other hand, Ref. \cite{6517345} considers the BTR strategy where only the timer is used. The waiting-time distribution and the power efficiency are derived in Ref. \cite{6517345} based on a recursive relation between the waiting times of two consecutive frames.

The general case that includes both the timer and the counter was investigated in \cite{Kim2013350,6078914,5743059,Alonso20122456,6279526}. To simplify the analysis, Ref. \cite{6078914,5743059,Alonso20122456,6279526} consider two extreme regimes, low-load scenario and high-load scenario, based on the observation that the timer expires before the counter reaches threshold $N$ if the traffic load is sufficiently low; otherwise, the counter triggers the Wakeup if the traffic load is high enough. The main drawback of these efforts is that they cannot obtain a unified expression of the power consumption and the mean delay over the whole traffic load regime. To fix this problem, Ref. \cite{Kim2013350} develops a model based on the renewal theory. In this paper, a sleep period followed by a busy period is considered as a renewal cycle. Though this paper assumes that frames will not arrive during the Sleep period, the derivation is still very complicated such that the results cannot provide a clear physical insight of EEE systems.

\subsection{Our Approach and Contributions}
In this paper, we analyze the BTR strategy of 10Gbase-T Ethernet, in which the two transmission directions are independent of each other. Our goal is to develop a unified model to predict the performance of the EEE protocol such that we can provide rules to select the proper values of parameter $\tau$ and $N$.

The BTR strategy is modeled as an M/G/1 queue with vacation time, which is governed by the frame arrival process via the timer threshold $\tau$ and the counter threshold $N$. First, we show that the classical M/G/1 queue with vacation time \cite{bertsekas1992data} cannot be directly applied to delineate the EEE protocol. We then develop a new approach to analyze the performance of the EEE protocol. Our key idea is to establish the connection between the vacation time and the arrival process to account for their dependency. We first derive the distribution of the number of arrivals during a vacation time based on an event tree of the BTR strategy, from which we obtain the mean vacation time and the power efficiency. Next, based on the number of arrivals at the end of the vacation, we derive a generalized P-K formula of the mean delay for EEE systems.

For a fixed traffic load, we find that the power efficiency converges to a constant while the mean delay is unbounded when $\tau$ and $N$ increase, which means that $\tau$ and $N$ should not be too large in the practical application of BTR strategy. Our results clearly show that the timer and the counter are compensating for each other and play different roles in different traffic rate regions. The timer $\tau$ bounds the delay incurred by the vacation when the traffic rate is smaller than $(N-1)/\tau$, while the counter $N$ limits the queue length during the vacation time when the load is larger than $(N-1)/\tau$. Therefore, for a given arrival rate $\lambda$, a proper choice of the parameters $\tau$ and $N$ should satisfy the condition $(N-1)/\tau=\lambda$. Our specific contributions are summarized as follows:
\begin{enumerate}[{1.}]
\item
We develop a new approach to analyze the M/G/1 queue with the vacation time that is governed by the arrival process and the parameters $\tau$ and $N$ of the BTR strategy.
\item
We derive a generalized P-K formula of mean delay for the M/G/1 queue with the vacation time controlled by the arrival process. We show that the classical P-K formula of mean delay is only a special case when the vacation time is independent of the arrival process.
\item
We show that the $\tau$ policy and $N$ policy of the BTR strategy are compensating each other. Our analysis demonstrates the impacts of parameters $\tau$ and $N$ on the power efficiency and mean delay. These results provide the rules to select the appropriate $\tau$ and $N$.
\end{enumerate}

The rest of this paper is organized as follows. In section II, we propose the vacation model which establishes the connection between the vacation time and the arrival process and derives the power efficiency of BTR strategy. In section III, we derive a generalized P-K formula of the mean delay which reduces to the traditional P-K formula when the vacation time is independent of the arrival process. In section IV, our model is validated by simulations. Furthermore, based on our analysis, we summarize two rules to select appropriate values of parameters. Section V concludes this paper.

\begin{figure*}[ht]
\centering
\includegraphics[scale=0.52]{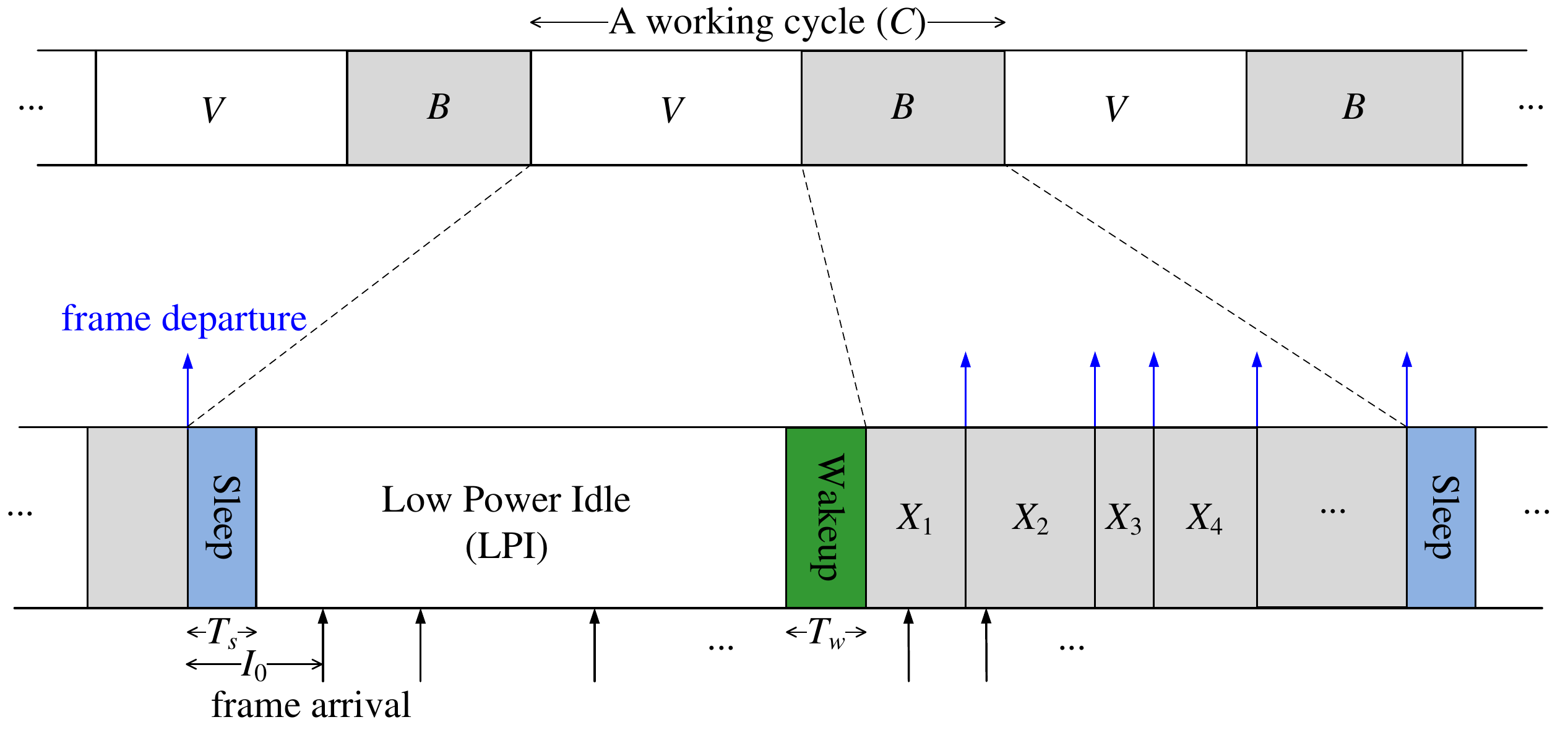}
\caption{The burst transmission (BTR) strategy of the 10GBase-T EEE protocol.}
\label{EEEpro}
\end{figure*}

The notations used in this paper are defined as follows:
\begin{itemize}
\item Parameters of the EEE protocol
\begin{description}[\IEEEsetlabelwidth{$C_2$}]
  \item[$T_s$] Duration of the Sleep
  \item[$T_w$] Duration of the Wakeup
  \item[$N$] Counter threshold
  \item[$\tau$] Timer threshold
  \item[$\varphi_0$] Power consumption (W per unit time) in LPI state
  \item[$\varphi_1$] Power consumption (W per unit time) in the busy period, Sleep and Wakeup
  \item[$\eta$] Power efficiency
\end{description}

\item System parameters
\begin{description}[\IEEEsetlabelwidth{$C_2$}]
  \item[$\lambda$] Frame arrival rate
  \item[$\mu$] Frame service rate
  \item[$D$] Mean delay of a frame
\end{description}

\item Variables in vacation model
\begin{description}[\IEEEsetlabelwidth{$C_2$}]
  \item[$a_n$] Probability that there are $n$ arrivals during the Sleep and LPI periods
  \item[$b_n$] Probability that there are $n$ arrivals during the Wakeup period
  \item[$h_n$] Probability that there are $n$ packets in the buffer at the end of vacation
  \item[$V$] Duration of a vacation
  \item[$C$] Duration of a cycle
  \item[$R_i$] Residual time for completing a vacation or a service seen by an arrival
  \item[$X_j$] Transmission (service) time of the $j$th frame
\end{description}
\end{itemize}

\begin{figure*} [ht]
\centering
\includegraphics[scale=0.64]{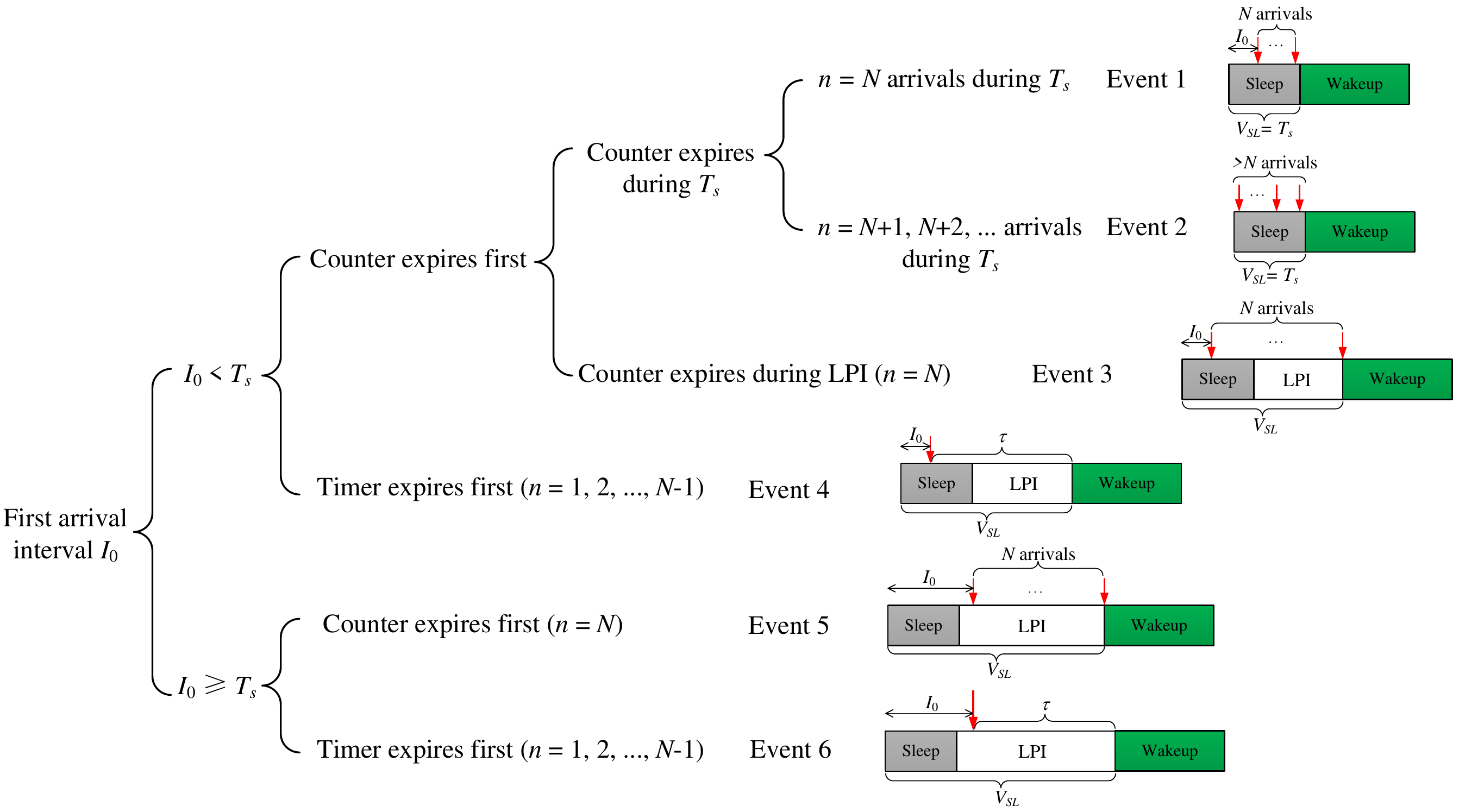}
\caption{Arrival event tree during $V_{SL}$.}
\label{Events}
\end{figure*}

\section{Vacation Model of EEE Protocol}

The process of the BTR strategy implemented on a 10GBase-T EEE interface in one direction of a link is a sequence of cycles, as Fig. \ref{EEEpro} shows, each of which consists of a vacation period followed by a busy period. The cycle, vacation period, and busy period are respectively denoted by $C$, $V$ and $B$. Each cycle begins when the transmission buffer becomes empty and the interface enters a vacation period. First, the interface turns off its transmitter through a Sleep operation, during which the interface is still in the high power mode, and the power consumption is denoted as $\varphi_h$. The duration time of the Sleep operation is 2.88$\mu s$, denoted as $T_s$. As specified in the 10GBase-T EEE protocol, the Sleep operation is uninterruptible. After the Sleep operation, the interface goes into an LPI (low power idle) mode, in which the power consumption is only $\varphi_l=0.1\varphi_h$. During the vacation period, the system initializes a timer and a counter when the first frame arrives, and the counter increases by one for each new arrival. As soon as the timer reaches the threshold $\tau$ or if there are $N$ frames waiting in the buffer, the vacation period terminates and the interface turns on its transmitter through a Wakeup operation. Similar to the Sleep operation, the Wakeup operation lasts a constant duration, 4.48$\mu s$, denoted as $T_w$, with the same power consumption $\varphi_h$. After the Wakeup operation, the interface starts the busy period, during which the frames are transmitted until the buffer becomes empty.

Depending on the selections of parameters $\tau$ and $N$, the BTR strategy can be further divided into several policies. If $\tau\rightarrow\infty$, the BTR strategy is referred to as $N$ policy \cite{yadin1963queueing}, which wakes up the interface by the counter threshold $N$. Similarly, if $N\rightarrow\infty$, the BTR strategy is called $\tau$ policy \cite{heyman1977t}, which wakes up the interface by the timer threshold $\tau$. The BTR strategy is called $\tau\&N$ policy as both $\tau$ and $N$ are finite. The basic idea of this policy is to guarantee that the delay and the backlogged queue length are both bounded.

It is clear that the interface is a queueing system with vacation time; the interface goes to sleep (or takes a vacation) once the buffer becomes empty. However, unlike the classical queueing model with vacation time, where the vacation time and the arrival process are statistically independent, the vacation time of the Ethernet interface with the BTR strategy is completely governed by the frame arrival process through $\tau$ and $N$ policies.

As the vacation time is determined by the number of arrivals during the vacation period, which is regulated by timer $\tau$ and counter $N$, our analysis of the BTR strategy starts with the derivation of the distribution of the number of arrivals during the vacation time, from which we derive the mean vacation time, the power efficiency, and the queuing delay. In modeling the BTR strategy, we adopt the following assumptions:
\begin{itemize}
\item The input traffic is a Poisson process with arrival rate $\lambda$;
\item The interface transmits the frames in a first in first out (FIFO) manner;
\item The frame transmission times $X_1,X_2,\dots$ are independent and identically distributed random variables with first and second moments $\overline{X}$ and $\overline{X^{2}}$, respectively;
\item The threshold $\tau$ is larger than the duration of Sleep $T_s$, because $T_s$ is very short and the timer $\tau\leq T_s$ is meaningless.
\end{itemize}

\begin{figure*} [hb]
\hrulefill
\begin{align}
a_n&=Pr\{n\text{ arrivals during}\ V_{SL}\} =
\left\{
\begin{array}{ll}
0, &          n=0\\
e^{-\lambda\tau}\frac{(\lambda\tau)^{n-1}}{(n-1)!},      & n=1,2,\dots,N-1\\
\sum_{n=0}^N e^{-\lambda T_s}\frac{(\lambda T_s)^{n}}{n!}- \sum_{n=1}^{N-1} e^{-\lambda\tau}\frac{(\lambda\tau)^{n-1}}{(n-1)!}       & n=N\\
e^{-\lambda T_s}\frac{(\lambda T_s)^{n}}{n!},            & n=N+1,N+2,\dots
\end{array}
\right.
\end{align}
\end{figure*}

\subsection{Vacation Model}
As we mentioned above, the key to model the BTR strategy is the number of arrivals during vacation period $V$, whose distribution is defined as follows:
$$h_n=Pr\{n\text{ arrivals during a vacation period}\ V\}.$$
The vacation period is divided into two independent parts. The first part, denoted as $V_{SL}$, consists of the Sleep and the LPI periods, while the second part is the Wakeup $T_w$. Accordingly, the number of arrivals during vacation period $V$ is the sum of the number of arrivals during $V_{SL}$ and that during $T_w$.

The tree depicted in Fig. \ref{Events} describes the complete arrival events that occurred during $V_{SL}$. We show in the following lemma that the distribution of the number of arrivals during $V_{SL}$ is determined by parameters $\tau$, $N$, and arrival rate $\lambda$.

\newtheorem{Lemma1}{Lemma}
\begin{Lemma1}
The distribution of the number of arrivals during $V_{SL}$ is given by equation (1).

\end{Lemma1}

\begin{IEEEproof}
Let $I_0$ be the time interval from the beginning of the vacation to the time when the first frame arrives, and $I_i$ be the inter-arrival time between the $i$th arrival and the $(i+1)$th arrival during the vacation, where $i=1,2,\dots.$ As Fig. \ref{Events} shows, there are six mutually exclusive arrival events that could occur during $V_{SL}$. They can be classified into the following four cases:
\begin{enumerate}[{1.}]
\item $n=0$

According to the BTR strategy, it is impossible that the link wakes up without any arrival, therefore
\begin{equation}
a_0=0,
\end{equation}
which implies that there is at least one arrival during $V_{SL}$.
\item $n=1,2,\dots,N-1$

Regardless if the first frame arrives before or after the end of the Sleep period, that is, $I_0\leq T_s$ or $I_0>T_s$, if the $V_{SL}$ terminates when the timer expires, then the number of arrivals must be less than $N$. In Fig. \ref{Events}, the timer expires in Events 4 and 6, thus we have
\begin{align}
a_n=&Pr\{I_0\leq T_s,n-1\ \text{arrivals in an interval}\ \tau\} \nonumber\\
      &+Pr\{I_0>T_s,n-1\ \text{arrivals in an interval}\ \tau\}\nonumber\\
    =&Pr\{n-1\ \text{arrivals in an interval}\ \tau\}\nonumber\\
    =&e^{-\lambda\tau\frac{(\lambda\tau)^{n-1}}{(n-1)!}},
\end{align}

for $n=1,2,\dots,N-1$.

\item $n=N+1,N+2,\dots$

The number of arrivals $n$ during $V_{SL}$  can be larger than $N$ only if they all arrived during Sleep period $T_s$, which occurs in Event 2. It follows that
\begin{equation}
a_n=Pr\{n\text{ arrivals in the interval}\ T_s\}=e^{-\lambda T_s}\frac{(\lambda T_s)^n}{n!},
\end{equation}
for $n=N+1,N+2,\dots$.
\item $n=N$

In Events 1, 3 and 5 shown in Fig. \ref{Events}, $V_{SL}$ is ended with exactly $N$ frames in the buffer. In Event 1, there are $N$ arrivals during Sleep period $T_s$, and the counter triggers the Wakeup immediately at the end of Sleep period. In Event 3, the first frame arrives during Sleep period $T_s$, that is, $I_0<T_s$, and $N$ frames arrive after the Sleep but before $I_0+\tau$. In Event 5, the first frame arrives after $T_s$, that is, $I_0>T_s$, and $N$ frames arrive before $I_0+\tau$. Given $\sum_{n=0}^\infty a_n=1$ , from (3) and (4), we have
\begin{align}
a_N=&Pr\{N\ \text{arrivals in the interval}\ T_s\}\nonumber \\
&+Pr\{I_0\leq T_s,T_s-T_0<I_1+\dots+I_{N-1}\leq\tau \}\nonumber \\
&+Pr\{I_0> T_s,I_1+\dots+I_{N-1}\leq\tau\}\nonumber \\
=&1-\sum_{n=1}^{N-1} e^{-\lambda\tau} \frac{(\lambda\tau)^{n-1}}{(n-1)!}-\sum_{n=N+1}^\infty e^{-\lambda T_s}\frac{(\lambda T_s)^n}{n!}\nonumber \\
=&\sum_{n=0}^N {e^{-\lambda T_s}\frac{(\lambda T_s)^n}{n!}}-\sum_{n=1}^{N-1} {e^{-\lambda\tau} \frac{(\lambda\tau)^{n-1}}{(n-1)!}}.
\end{align}
\end{enumerate}

\end{IEEEproof}

Let $b_n$ be the probability that there are $n$ arrivals during Wakeup period $T_w$, we have
\begin{equation}
b_n=Pr\{n\text{ arrivals during}\ T_w\}=e^{-\lambda T_w}\frac{(\lambda T_w)^n}{n!},
\end{equation}
for $n=0,1,\dots.$ From Lemma 1 and (6), the generating functions of $a_n$ and $b_n$ are respectively given as follows.
\begin{align}
A(z)=&\sum_{n=0}^\infty a_nz^n\nonumber \\
=&\sum_{n=0}^{N-1}{e^{-\lambda T_s}\frac{(\lambda T_s)^n}{n!}}\big(z^N-z^n\big)\nonumber \\
&-\sum_{n=0}^{N-2}{e^{-\lambda \tau}\frac{(\lambda \tau)^n}{n!}}\big(z^N-z^{n+1}\big)+e^{-\lambda T_s(1-z)}
\end{align}
and
\begin{align}
B(z)=\sum_{n=0}^\infty b_nz^n=e^{-\lambda T_w}\sum_{n=0}^\infty {\frac{(\lambda T_w z)^n}{n!}}=e^{-\lambda T_w(1-z)}
\end{align}
Since the number of arrivals during vacation period $V$ is the sum of arrivals during $V_{SL}$ and that of $T_w$, the probability that there are $n$ arrivals during vacation period $V$ is the convolution of $a_n$ and $b_n$ given as follows:
\begin{equation}
h_n=\sum_{k=0}^n a_{n-k}b_k
\end{equation}
Thus, the generating function of $h_n$ is given by $H(z)=A(z)B(z)$, and we immediately obtain the mean number of arrivals during a vacation period as follows:
\begin{align}
    \overline{\alpha}=&H'(1)\nonumber \\
    =&\lambda(T_w+T_s)+\sum_{n=0}^{N-1}e^{-\lambda T_s}\frac{(\lambda T_s )^n}{n!}(N-n)\nonumber \\
    &-\sum_{n=0}^{N-2} e^{-\lambda \tau}\frac{(\lambda\tau)^n}{n!}[N-(n+1)].
\end{align}
Furthermore, it follows from Little's Law that the mean vacation time under the $\tau\&N$ policy is given by
\begin{align}
\overline{V}=&\frac{\overline{\alpha}}{\lambda}=\frac{H'(1)}{\lambda}\nonumber \\
=&T_s+\sum_{n=0}^{N-1}e^{-\lambda T_s}\frac{(\lambda T_s )^n}{n!}\bigg(\frac{N-n}{\lambda}\bigg)\nonumber \\
&-\sum_{n=0}^{N-2}e^{-\lambda \tau}\frac{(\lambda\tau)^n}{n!}\bigg(\frac{N-n-1}{\lambda}\bigg)+T_w.
\end{align}
Recall that the $\tau$ policy is a $\tau\&N$ policy with $N\rightarrow\infty$ and the $N$ policy is a $\tau\&N$ policy with $\tau\rightarrow\infty$. Thus, letting $N\rightarrow\infty$, the mean vacation time under the $\tau$ policy can be derived from (11) as follows:
\begin{equation}
\overline{V}_\tau=\lim_{N\rightarrow\infty}\overline{V}=\frac{1}{\lambda}+\tau+T_w,
\end{equation}
which includes three components: (1) the expected time from the beginning of the vacation period to the first arrival time, (2) the duration of the timer, and (3) the Wakeup period. Similarly, $\tau\rightarrow\infty$ results in the mean vacation time under the $N$ policy:
\begin{equation}
\overline{V}_N=\lim_{\tau\rightarrow\infty}\overline{V}=T_s+\sum_{n=0}^{N-1}e^{-\lambda T_s}\frac{(\lambda T_s )^n}{n!}\bigg(\frac{N-n}{\lambda}\bigg)+T_w.
\end{equation}
Since the Sleep period $T_s$ is very small, we can ignore the probability that there are at least $N-1$ arrivals during $T_s$, that is, $\sum_{n=N-1}^\infty e^{-\lambda T_s}\frac{(\lambda T_s )^n}{n!}\approx0,$ then we have:
\begin{align}
\overline{V}_N=&T_s+\frac{N}{\lambda}\sum_{n=0}^{N-1}e^{-\lambda T_s}\frac{(\lambda T_s )^n}{n!}+T_w\nonumber \\
&-T_s\sum_{n=0}^{N-2}e^{-\lambda T_s}\frac{(\lambda T_s )^n}{n!} \approx\frac{N}{\lambda}+T_w,
\end{align}
which indicates that the Wakeup is triggered by the $N$th arrival in the $N$ policy. Equations (11) through (14) demonstrate that the mean vacation time is determined by timer $\tau$ and counter $N$. In particular, $\overline{V}_\tau$ and $\overline{V}_N$ are proportional to $\tau$ and $N $, respectively.

\subsection{Power Efficiency}
The design of the EEE protocol aims to maximize power efficiency, which is defined as the percentage of energy the EEE protocol can save. As we previously mentioned in this section, the power consumption during LPI periods is $\varphi_l$ and that during Sleeps, Wakeups, and busy periods it is $\varphi_h$, and $\varphi_l=10\%\varphi_h$. Let $\rho$ be the offered load, that is, $\rho=\lambda \overline{X}$, then the mean cycle time is given by $\overline{C}=\overline{V}/(1-\rho)$. Because the mean time of the LPI period in a cycle is $\overline{V}-T_s-T_w$, the mean power consumption of the EEE system is given by:
\begin{equation}
\varphi_{EEE}=\frac{(\overline{V}-T_s-T_w)\varphi_l+(\overline{C}-\overline{V}+T_s+T_w)\varphi_h}{\overline{C}}.
\end{equation}
Without the EEE protocol, power consumption of an interface is a constant $\varphi_h$. It follows that the power efficiency is given by
\begin{align}
\eta=&\frac{\varphi_{h}-\varphi_{EEE}}{\varphi_{h}}\nonumber \\
=&\frac{(\overline{V}-T_s-T_w)(\varphi_h-\varphi_l)}{\overline{V}/(1-\rho)\varphi_h}\nonumber \\
=&\bigg(1-\frac{T_s+T_w}{\overline{V}}\bigg)\cdot\frac{(1-\rho)(\varphi_h-\varphi_l)}{\varphi_h}.
\end{align}

This expression clearly demonstrates that power efficiency $\eta$ increases with $\overline{V}$ for a given traffic load $\rho$. Moreover, when $n\rightarrow\infty$, we can obtain the power efficiency of the $\tau$ policy:
\begin{equation}
\eta_\tau=\lim_{N\rightarrow\infty}\eta=\bigg(1-\frac{T_s+T_w}{\frac{1}{\lambda}+\tau+T_w}\bigg)
\cdot\frac{(1-\rho)(\varphi_h-\varphi_l)}{\varphi_h}.
\end{equation}
Similarly, when $\tau\rightarrow\infty$, we can obtain the following power efficiency of the $N$ policy:
\begin{align}
\eta_N=&\lim_{\tau\rightarrow\infty}\eta \nonumber \\
=&\bigg(1-\frac{T_s+T_w}{T_s+\sum_{n=0}^{N-1}e^{-\lambda T_s}\frac{(\lambda T_s )^n}{n!}(\frac{N-n}{\lambda})+T_w}\bigg)\nonumber \\
&\cdot\frac{(1-\rho)(\varphi_h-\varphi_l)}{\varphi_h}\nonumber \\
\approx&\bigg(1-\frac{T_s+T_w}{\frac{N}{\lambda}+T_w}\bigg)\cdot\frac{(1-\rho)(\varphi_h-\varphi_l)}{\varphi_h},
\end{align}
where the approximation is obtained by ignoring the probability that there are at least $N-1$ arrivals during $T_s$.

\newcounter{TempEqCnt}
\setcounter{TempEqCnt}{\value{equation}}
\setcounter{equation}{33}

\begin{figure*} [hb]
\hrulefill
\begin{align}
D_N=&\frac{\lambda\overline{X^{2}}}{2(1-\rho)}+\frac{(\lambda T_w+\lambda T_s)^2+e^{-\lambda T_s}\sum_{k=0}^{N-1}\frac{(\lambda T_s )^k}{k!}[2\lambda T_w(N-k)+N(N-1)-k(k-1)]}
{2\lambda[\lambda(T_w+T_s)+e^{-\lambda T_s}\sum_{k=0}^{N-1}\frac{(\lambda T_s )^k}{k!}(N-k)]}+\overline{X}\nonumber \\
\approx &\frac{\lambda\overline{X^{2}}}{2(1-\rho)}+\frac{(N+\lambda T_w)^2-N}{2\lambda(N+\lambda T_w)}+\overline{X}.
\end{align}
\end{figure*}
\setcounter{equation}{\value{TempEqCnt}}

\section{P-K Formula of Mean Delay}
The mean delay of a traditional M/G/1 queueing system with vacation is given by the following well-known P-K formula:
\begin{equation}
D=\frac{\lambda\overline{X^{2}}}{2(1-\rho)}+\frac{\overline{V^{2}}}{2\overline{V}}+\overline{X},
\end{equation}
where $\overline{V^{2}}$ is the second moment of the vacation time. Ref. \cite{bertsekas1992data} states that this formula can be applied to different scenarios and points out ``{the length of a vacation interval need not be independent of the customer¡¯s arrival and service times.'' If this is the case, the mean delay can be readily obtained once the distribution of the vacation time is determined. In an EEE system, however, the vacation time is completely governed by the arrival process. In contrast to the above comment on equation (19), the following two counter examples show that the above P-K formula is invalid in a system controlled by the EEE protocol:
\begin{enumerate}[{1.}]
\item In the Appendix, we show that the following classical relationship between the number of arrivals in a vacation time and the distribution of vacation time $V$ \cite{kleinrock1975theory} does not hold in EEE systems:
\begin{equation}
H(z)=V^*(\lambda-\lambda z),
\end{equation}
where $V^* (s)$ is the generating function of the vacation time distribution.
\item The counter example provided in the Appendix shows that (19) is invalid for EEE systems even if the exact vacation time distribution is known.
\end{enumerate}

As we explain in the Appendix, both of these two counter examples are induced by the dependency between the vacation time and the arrival process. In this subsection, we derive the appropriate mean delay formula for EEE systems.

\begin{figure}
\centering
\includegraphics[scale=0.865]{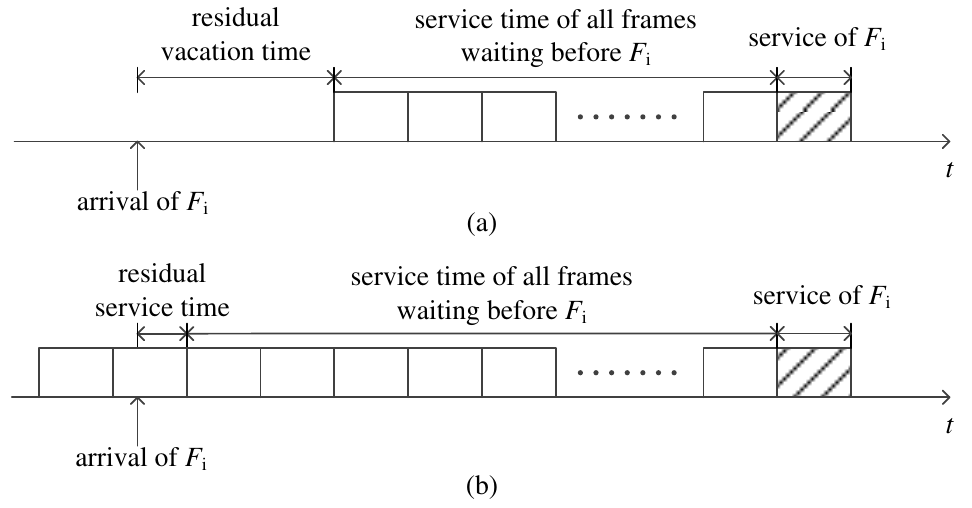}
\caption{ Frame $F_i$ arrives during (a) a vacation period (b) a busy period.}
\label{Resitime}
\end{figure}

Consider the $i$th arrival frame, denoted as $F_i$, in cycle $C$. Frame $F_i$ may arrive during the busy period of $C$ or the vacation period of $C$. Before frame $F_i$ receives service, it waits in the buffer for the completion of the current service period or current vacation period, and then for the services of all frames waiting in the buffer ahead of $F_i$. Let $R_i$ be the residual time, either service or vacation, of frame $F_i$. Following the same argument Ref. \cite{bertsekas1992data} states, we have:
\begin{equation}
D=\frac{\overline{R}}{1-\rho}+\overline{X},
\end{equation}
in which the mean residual time $\overline{R}=E[R_i]$ is the key to derive the P-K formula. As we explain in the Appendix, similar to the vacation time, the residual vacation time is related to the arrival process. Let $V_n$ denote a vacation period terminated with $n$ arrivals, and $P_n$ be the probability that frame $F_i$ arrives in a $V_n$. We need the following lemma to derive the P-K formula.

\newtheorem{Lemma2}[Lemma1]{Lemma}
\begin{Lemma2}
If a frame arrives during a vacation period, then the probability that the frame arrives in a $V_n$ is given by:
\begin{equation}
P_n=\frac{nh_n}{H'(1)}.
\end{equation}
\end{Lemma2}

\begin{IEEEproof}
Suppose the EEE system undergoes $k_n$ vacation periods $V_n$, for $n=1,2,\dots,$ in the time interval $[0,T]$. Then the probability $h_n$  that a vacation period is a $V_n$ is defined by:
\begin{equation}
h_n=\lim_{T\rightarrow\infty}\frac{k_n}{\sum_{n=1}^\infty k_n}.
\end{equation}
Thus, the conditional probability $P_n$ that a frame arrives in a $V_n$ is given by:
\begin{equation}
P_n=\lim_{T\rightarrow\infty}\frac{nk_n}{\sum_{n=1}^\infty nk_n}=\frac{nh_n}{\sum_{n=1}^\infty nh_n}=\frac{nh_n}{H'(1)}.
\end{equation}
\end{IEEEproof}
For a given traffic load $\rho$, the following theorem shows that the mean delay of an EEE system is completely governed by $h_n$, which is determined by the BTR strategy of EEE protocols.

\newtheorem{Theorem1}{Theorem}
\begin{Theorem1}
The mean delay of EEE systems is given by:
\begin{equation}
D=\frac{\lambda\overline{X^{2}}}{2(1-\rho)}+\frac{H''(1)}{2\lambda H'(1)}+\overline{X}.
\end{equation}
\end{Theorem1}
\begin{IEEEproof}
As Fig. \ref{Resitime} illustrates, a frame $F_i$ may experience a residual time waiting for the completion of a service or a vacation, depending on whether it arrives during a busy period or a vacation period. Define the following indicator variable:
\begin{align}
\xi=
\left\{
\begin{array}{ll}
0, & \text{if a frame arrives during a vacation period}\\
1, & \text{if a frame arrives during a busy period},
\end{array}
\right.\nonumber
\end{align}
The mean residual time can then be expressed as follows:
\begin{align}
\overline{R}&=E[R_i|\xi=0]\times Pr\{\xi=0\}+E[R_i|\xi=1]\times Pr\{\xi=1\}\nonumber \\
&=E[R_i|\xi=0]\times (1-\rho)+E[R_i|\xi=1]\times \rho.
\end{align}
The conditional expectation $E[R_i|\xi=1]$ in (26) can be solved by using the graphic method described in \cite{bertsekas1992data} and expressed as follows:
\begin{equation}
E[R_i|\xi=1]=\frac{1}{2\rho}\lambda\overline{X^{2}}.
\end{equation}
The mean residual vacation time can be obtained from the following expression of the conditional expected value:
\begin{align}
E[&R_i|\xi=0]=\sum_{n=1}^\infty E[R_i|\xi=0,\text{frame }i\text{ arrives in a }V_n]P_n.
\end{align}
Let $Q_i$ be the number of frames that arrive during time period $R_i$, or equivalently, the queue length behind the $i$th frame in the buffer when the vacation period terminates. Applying Little's Law to (28), we have:
\begin{align}
\lambda E[R_i|\xi=0]&=\sum_{n=1}^\infty \lambda E[R_i|\xi=0,\text{frame }i\text{ arrives in a }V_n] P_n  \nonumber \\
&=\sum_{n=1}^\infty E[Q_i|\xi=0,\text{frame }i\text{ arrives in a }V_n] P_n\nonumber \\
&=\sum_{n=1}^\infty\bigg[\frac{(n-1)+(n-2)+\dots+1+0}{n}\bigg]\frac{nh_n}{H'(1)}\nonumber \\
&=\frac{1}{2H'(1)}\sum_{n=1}^\infty n(n-1)h_n \nonumber \\
&=\frac{H''(1)}{2H'(1)}.
\end{align}
The theorem is established by combining (21), (26), (27), and (29).
\end{IEEEproof}
If the vacation time is independent of the arrival process in an M/G/1 queue with vacations, then the relation $H(z)=V^*(\lambda-\lambda z)$ holds. We have
\begin{equation}
H'(1)={V^{*}}'(\lambda-\lambda z)|_{z=1}=\lambda\overline{V}
\end{equation}
and
\begin{equation}
H''(1)={V^*}''(\lambda-\lambda z)|_{z=1}=\lambda^2\overline{V^2}.
\end{equation}
Substituting (30) and (31) into (25), we obtain the classical P-K formula of mean delay (19) again. Thus, the formula (25) given in Theorem 1 is more general than the classical P-K formula (19). The later is a special case of (25) when the relation $H(z)=V^*(\lambda-\lambda z)$ holds.

Substituting the generating function $H(z)$ into (25), we can obtain the mean delay of the $\tau\&N$ policy:
\begin{equation}
D=\frac{\lambda\overline{X^{2}}}{2(1-\rho)}+\frac{A}{2 \lambda B}+\overline{X}
\end{equation}
where
\begin{align}
A=&(\lambda T_w+\lambda T_s)^2+\sum_{k=0}^{N-1}\bigg\{e^{-\lambda T_s}\frac{(\lambda T_s )^k}{k!}[2\lambda T_w(N-k)\nonumber \\
&+N(N-1)-k(k-1)]\bigg\} -\sum_{k=0}^{N-2}\bigg\{e^{-\lambda \tau}\frac{(\lambda \tau)^k}{k!}\nonumber \\
&\times[2\lambda T_w(N-k-1) +N(N-1)-k(k+1)]\bigg\}\nonumber
\end{align}
and
\begin{align}
B=&\lambda(T_w+T_s)+\sum_{k=0}^{N-1}e^{-\lambda T_s}\frac{(\lambda T_s )^k}{k!}(N-k)\nonumber \\
&-\sum_{k=0}^{N-2}e^{-\lambda \tau}\frac{(\lambda \tau)^k}{k!}(N-k-1).\nonumber
\end{align}
Letting $N\rightarrow\infty$, we obtain the mean delay of the $\tau$ policy:
\begin{equation}
D_\tau=\frac{\lambda\overline{X^{2}}}{2(1-\rho)}+\frac{[\lambda(\tau+T_w)]^2
+2\lambda(\tau+T_w)}{2\lambda[1+\lambda(\tau+T_w)]}+\overline{X}.
\end{equation}
Similarly, $\tau\rightarrow\infty$ yields the mean delay of the $N$ policy which is shown in equation (34) where the approximation is obtained by ignoring the probability that more than $N-1$ frames arrive within the Sleep period $T_s$.


\section{Power Efficiency and Delay Tradeoff}
In this section, we study the performance of EEE systems in terms of power efficiency and mean delay through the results obtained in Section II and Section III. We demonstrate the tradeoff between different parameters, which in turn provides the principle to optimize the design of the EEE protocol. Our goal is to find the rules to select appropriate values for parameters $\tau$ and $N$ with regards to the performances of the EEE protocol. In the following analysis, we assume that the average frame size is 1250 bytes, and thus the average frame transmission rate is 1 frame/$\mu$s.
\subsection{Timer versus Counter}
The key parameters of the EEE protocol are timer threshold $\tau$ and counter threshold $N$. The analytical results presented in the previous section reveal that the vacation time is governed by the arrival process. Thus, the behavior of an EEE system can be characterized by the inter-relationship between these two parameters and frame arrival rate $\lambda$.

In the $\tau\&N$ policy, the timer and the counter are initialized upon the first arrival during the vacation. According to Little's Law, the mean vacation time $\overline{V}$ given by (11) can be approximately rewritten as follows:

\setcounter{equation}{34}
\begin{equation}
\overline{V}\approx\min\{{\overline{V}}_\tau,{\overline{V}}_N\}
=\min\bigg\{\frac{1}{\lambda}+\tau,\frac{N}{\lambda}\bigg\}+T_w.
\end{equation}
Therefore, depending on the frame arrival rate $\lambda$, the Wakeup could be triggered by the timer or the counter. The following are possible scenarios:
\begin{enumerate}[{1.}]
\item If the arrival rate is small, say $\lambda<\frac{N-1}{\tau}$, then $\overline{V}\approx\frac{1}{\lambda}+\tau+T_w$ implies that the Wakeup is most likely triggered by the timer. In this case, the $\tau \& N$ policy acts similar to the $\tau$ policy, which ensures that the residual vacation time of a frame arrived during the vacation period is bounded by $\tau+T_w$.
\item If the arrival rate is large, say $\lambda>\frac{N-1}{\tau}$, then $\overline{V}\approx\frac{N}{\lambda}+T_w$ implies that the Wakeup is mainly triggered by the counter. In this case, the $\tau \& N$ policy performs as the $N$ policy, such that the backlogged queue length at the end of vacation can be bounded. Therefore, the queueing delay induced by the vacation period can also be bounded.
\item For a moderate arrival rate $\lambda\approx\frac{N-1}{\tau}$, the Wakeup could be triggered by the timer or the counter with almost the same probability.
\end{enumerate}
Presumably, the instantaneous arrival rate $\lambda(t)$ may fluctuate around the steady state rate $\lambda=\lim_{t\rightarrow\infty}\lambda(t)$ over time $t$. The above property indicates that the $\tau \& N$ policy is able to adapt to the fluctuations of the arrival rate, and it will strike a balance between $\tau$ policy and $N$ policy. This property is particularly useful in the face of bursty arrival traffic.

\begin{figure}
\centering
\includegraphics[scale=0.65]{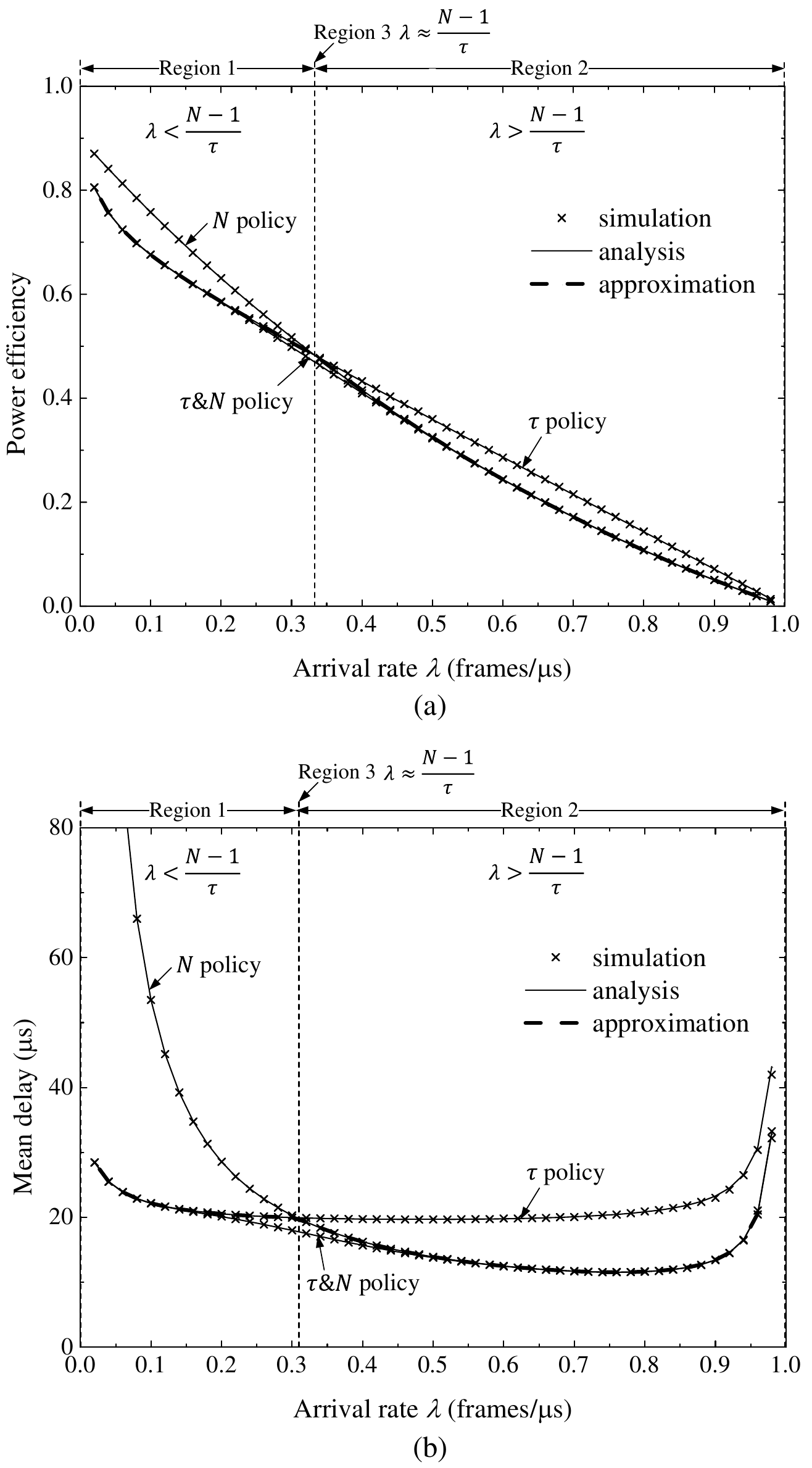}
\caption{ Performance evaluation: (a) power efficiency and (b) mean delay vs. $\lambda$ , where $\tau=30\mu s$ and $N=11$.}
\label{perfromI}
\end{figure}

The power efficiency and the mean delay versus the frame arrival rate $\lambda$, where $\tau=30\mu s$ and $N=11$, are depicted in Fig. \ref{perfromI} which plots (16), (17), and (18) in Fig. \ref{perfromI}(a), and (32), (33), and (34) in Fig. \ref{perfromI}(b). The simulation results perfectly agree with our analytical predictions in power efficiency and mean delay. Power efficiency $\eta$ is proportional with $\overline{V}$ as we explained in Section II-B. Therefore, similar to (35), the power efficiency $\eta$ can be approximately expressed as follows:
\begin{align}
\eta&\approx\min\{\eta_\tau,\eta_N\}\nonumber \\
&=\min\bigg\{1-\frac{T_s+T_w}{\frac{1}{\lambda}+\tau+T_w},1-\frac{T_s+T_w}{\frac{N}{\lambda}+T_w}\bigg\}
\frac{(1-\rho)(\varphi_h-\varphi_l)}{\varphi_h}.
\end{align}
Accordingly, we obtain the following approximation of the mean delay $D$ given by (32):

\begin{align}
D\approx&\min\{D_\tau,D_N\}\nonumber \\
=&\min\bigg\{\frac{[\lambda(\tau+T_w)]^2
+2\lambda(\tau+T_w)}{2\lambda[1+\lambda(\tau+T_w)]},\frac{(N+\lambda T_w)^2-N}{2\lambda(N+\lambda T_w)}\bigg\}\nonumber \\
&+\frac{\lambda\overline{X^{2}}}{2(1-\rho)}+\overline{X}.
\end{align}

As Fig. \ref{perfromI} shows, the approximations given by (36) and (37) are quite accurate for practical applications. Also, despite the power efficiencies $\eta$, $\eta_\tau$, and $\eta_N$ in three different regions are all close to each other, the discrepancy among the mean delays $D$, $D_\tau$, and $D_N$ can be quite substantial. For example, $D_N=54\mu s$ while $D_\tau=D=22\mu s$ when $\lambda=0.1$ frames/$\mu s$, and $D_\tau=21\mu s$ while $D_N=D=12\mu s$ when $\lambda=0.8$ frames/$\mu s$.

Since Ethernet traffic is typically bursty, the $\tau\&N$ policy adapts to traffic fluctuations. As we explained above, the $N$ policy suffers from a large delay in region $\lambda<(N-1)/\tau$, since $\lambda$ is small and it may take a long time for the counter to reach $N$. In contrast, the $\tau$ policy faces a large delay in region $\lambda>(N-1)/\tau$, because the large rate $\lambda$ incurs a large number of backlogged frames during time $\tau$. As a compromise, the $\tau\&N$ policy can avoid large delays in these two side regions if parameters $\tau$ and $N$ are selected according to the following rule:
\newtheorem{rule1}{EEE}
\begin{rule1}
For a given steady state traffic rate $\lambda$, the selection of parameters $\tau$ and $N$ should comply with the following condition:
\begin{equation}
\frac{N-1}{\tau}=\lambda.
\end{equation}
\label{rule1}
\end{rule1}

\subsection{Power Efficiency versus Mean Delay}
In this subsection, we investigate the power efficiency $\eta$ and the mean delay $D$ of the $\tau \&N$ policy under the assumption that (38) of the selection rule EEE \ref{rule1} is observed. For a given arrival rate $\lambda=1/3$ frames/$\mu s$, the power efficiency $\eta$ is plotted as a function of $N$ in Fig. \ref{perfromII}(a), from which we can see that $\eta$ increases with $N$ and finally converges to a constant. As the expression (16) demonstrates, the power efficiency $\eta$ increases with the mean vacation time $\overline{V}$, which increases with $\tau$ and $N$. In particular, if $\overline{V}$ goes to infinity with $\tau$ and $N$, the ratio $\frac{T_w+T_s}{\overline{V}}$ approaches 0, meaning that the overhead, including the Sleep time and the Wakeup time, is negligible if the LPI period is sufficiently long. In this case, the power efficiency reaches the following maximal value:
\begin{equation}
\eta^*=\lim_{\overline{V}\rightarrow\infty}=\frac{(1-\rho)(\varphi_h-\varphi_l)}{\varphi_h}
\end{equation}
The mean delay $D$ is almost linearly proportional to $\tau$ and $N$, as Fig. \ref{perfromII}(b) shows. Intuitively, if parameter $\tau$ and $N$ are too large, then a large number of frames that arrive during the vacation time would have to wait for a long time before the timer expires or the counter reaches the threshold $N$. Furthermore, the large number of backlogged frames will also affect the queueing delay of the frames.

\begin{figure}
\centering
\includegraphics[scale=1.25]{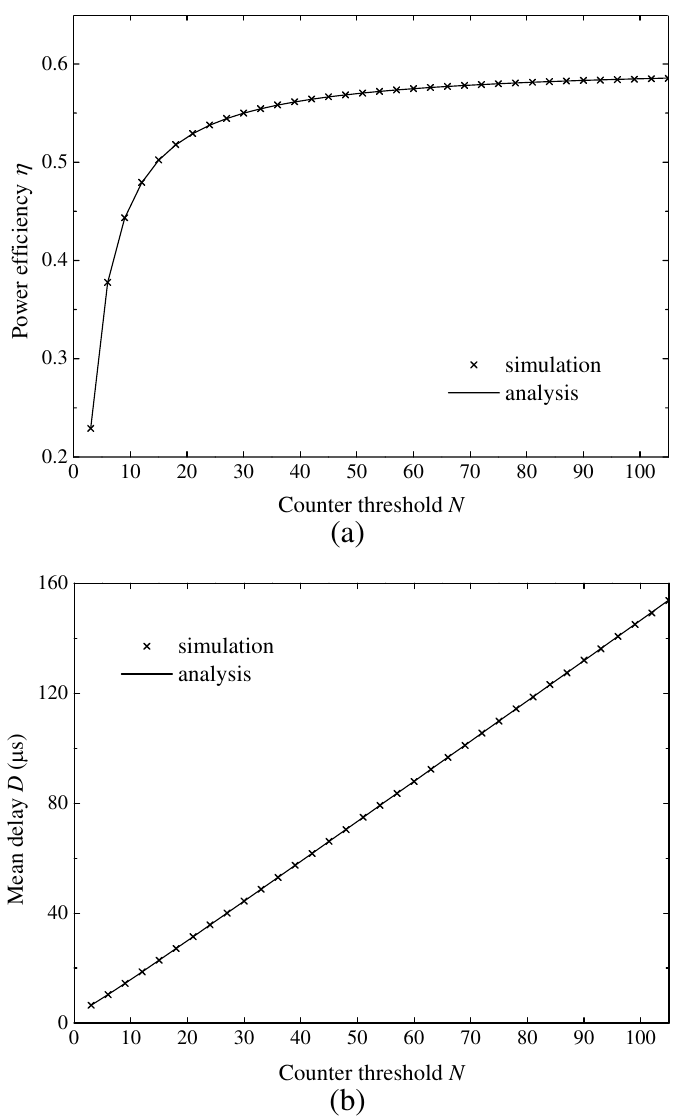}
\caption{Performance evaluation: (a) power efficiency and (b) mean delay versus $N$, where $\frac{N-1}{\tau}=\lambda$ and $\lambda=1/3$ frames/$\mu s$.}
\label{perfromII}
\end{figure}

\begin{figure}
\centering
\includegraphics[scale=0.455]{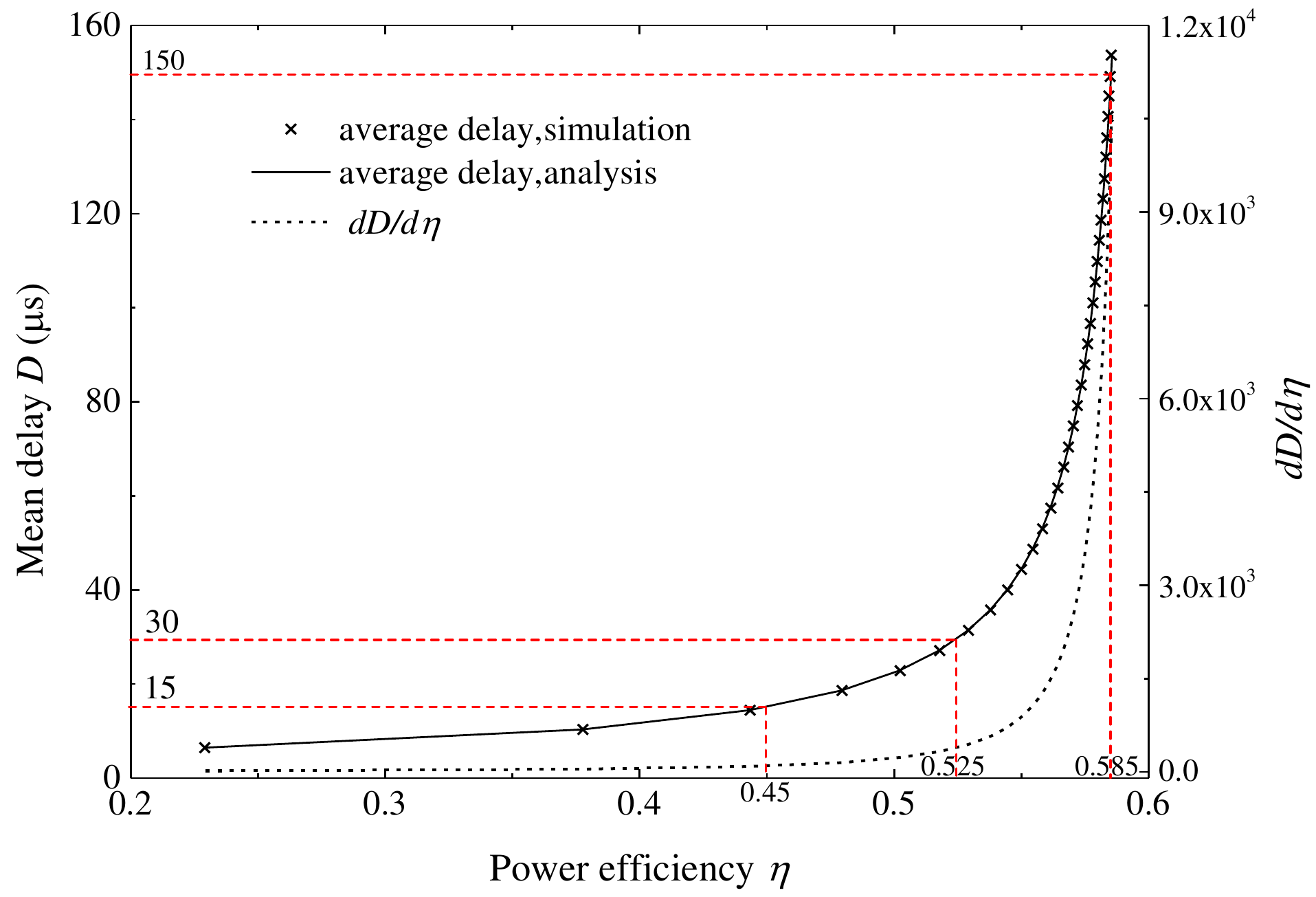}
\caption{Mean delay versus power efficiency when $\lambda=1/3$ frames/$\mu s$.}
\label{perfromIII}
\end{figure}

The results in Fig. \ref{perfromII} indicate that excessive large $\tau$ and $N$ degrade delay performance while marginally enhancing the power efficiency. Under condition $\frac{N-1}{\tau}=\lambda$ of the selection rule EEE 1, we obtain the following mean delay as a function of the power efficiency from (18) and (34):
\begin{equation}
D\approx\frac{\lambda\overline{X^{2}}}{2(1-\rho)}
+\frac{T_s+T_w}{2(1-\frac{\eta}{\eta^*})}-\frac{T_s+T_w\frac{\eta}{\eta^*}}{2\lambda(T_s+T_w)}
+\overline{X}.
\end{equation}
Taking the derivative of (40) with respect to $\eta$, we have:
\begin{equation}
\frac{dD}{d\eta}\approx\frac{T_s+T_w}{2\eta^*(1-\frac{\eta}{\eta^*})^2}
-\frac{T_w}{2\lambda \eta^*(T_s+T_w)}.
\end{equation}
Eq. (40) and Eq. (41) demonstrate that the power efficiency of the EEE protocol is improved at the expense of increasing mean delay. When power efficiency $\eta$ is small, it can be significantly improved with a small increase of delay $D$, while the delay $D$ skyrockets to infinity as $\eta$ approaches $\eta^*$. For example, as Fig. 6 shows, the mean delay $D$ is doubled when $\eta$ increases from 0.45 to 0.525, while it expands five times when $\eta$ only slightly increases from 0.525 to 0.585. Thus, our second rule to select the parameters of the EEE protocol is given as follows:
\newtheorem{rule2}[rule1]{EEE}
\begin{rule2}
 Parameter $N$ of the EEE protocol can be selected from (34) according to a given mean delay requirement $D$.
 \label{rule2}
\end{rule2}

According to the above two selection rules, counter threshold $N$ can be derived from Eq. (34) when the delay requirement is given. Then timer threshold $\tau$ is determined by rule EEE \ref{rule1} and the corresponding power efficiency is given by Eq. (40). Table I lists the parameter selection procedures of EEE systems under different arrival rates. Here, the delay requirements are expressed in terms of the mean delay of the EEE system with FTR strategy as a reference.

For the FTR strategy, the power efficiency and mean delay can be obtained from (16) and (25), respectively. The first and the second moments of $h_n$ are respectively given by
\begin{equation}
H'(1)=\lambda(T_s+T_w)+e^{-\lambda T_s},\nonumber
\end{equation}
and
\begin{equation}
H''(1)=[\lambda(T_s+T_w)]^2+2\lambda T_w e^{-\lambda T_s}.\nonumber
\end{equation}
Assuming the service rate is 1 frame/$\mu s$ for the FTR strategy, we obtain the following mean delay and power efficiency with respect to different traffic loads:
\begin{enumerate}[{(1)}]
\item  $\lambda_1=1/5\text{ frames/}\mu s, D_1=5.0261\mu s, \eta_1=0.1990;$
\item  $\lambda_2=1/3\text{ frames/}\mu s, D_2=5.0380\mu s, \eta_2=0.0810;$
\item  $\lambda_3=3/5\text{ frames/}\mu s, D_2=5.4609\mu s, \eta_3=0.0139.$
\end{enumerate}
\begin{table}[htbp]
\newcommand{\tabincell}[2]{\begin{tabular}{@{}#1@{}}#2\end{tabular}}
\caption{Parameters Selection under Different Arrival Rates and Delay Requirements}
\label{table1}
\centering
\begin{threeparttable}[b]
\begin{tabular}{|c|c|c|c|c|c|}
\hline
\tabincell{c}{ $\lambda$ \\(frames/$\mu s$)} & Strategy &
\tabincell{c}{ $D$ \\ ($\mu s$)} &  $\eta$ &  $\ \ N\ \ $  & \tabincell{c}{ $\ \ \tau\ \ $ \\($\mu s$)}\\
\hline
                    & FTR & $D_1$ & $\eta_1$ & 1 & 0 \\ \cline{2-6}
                    &     & $\ \ 1.23D_1\ \ $  &  $\ \ 1.53\eta_1\ \ $ & 2  & 5 \\ \cline{3-6}
                    &     & $2.03D_1$  &  $2.36\eta_1$ & 4  & 15 \\ \cline{3-6}
  $\lambda_1=1/5$   & BTR & $5.20D_1$  &  $3.12\eta_1$ & 11 & 50 \\ \cline{3-6}
                    &     & $9.91D_1$  &  $3.35\eta_1$ & 21 & 100 \\ \cline{3-6}
                    &     & $19.49D_1$ &  $3.48\eta_1$ & 41 & 200 \\ \cline{1-6}
\hline
                    & FTR & $D_2$ & $\eta_2$ & 1 & 0 \\ \cline{2-6}
                    &     & $1.09D_2$ & $1.76\eta_2$ & 2 & 3 \\ \cline{3-6}
                    &     & $2.05D_2$ & $4.66\eta_2$ & 6 & 15 \\ \cline{3-6}
  $\lambda_2=1/3$   & BTR & $5.09D_2$ & $6.33\eta_2$ & 17 & 48 \\ \cline{3-6}
                    &     & $10.23D_2$ & $6.88\eta_2$ & 35 & 102 \\ \cline{3-6}
                    &     & $20.06D_2$ & $7.14\eta_2$ & 69 & 204 \\ \cline{1-6}
\hline
                    & FTR & $D_3$ & $\eta_3$ & 1  &  0 \\ \cline{2-6}
                    &     & $1.01D_3$ & $1.62\eta_3$ & 2  & 1.67 \\ \cline{3-6}
                    &     & $1.99D_3$ & $15.96\eta_3$ & 10 & 15 \\ \cline{3-6}
  $\lambda_3=3/5$   & BTR & $5.04D_3$ & $22.27\eta_3$ & 31 & 50  \\ \cline{3-6}
                    &     & $9.92D_3$ & $24.13\eta_3$ & 65 & 105 \\ \cline{3-6}
                    &     & $20.23D_3$ & $25.03\eta_3$ & 133 & 220 \\ \cline{1-6}

\end{tabular}
\begin{tablenotes}
\item
$\lambda$: arrival rate, $D$: mean delay, $\eta$: power efficiency.
\end{tablenotes}
\end{threeparttable}
\end{table}

\section{Conclusion}
This paper presents an analytical model for the BTR strategy of 10GBase-T Energy Efficient Ethernet and derives the power efficiency and mean delay of the system. The proposed model is the first approach that can dispose of the dependency between the vacation time and the arrival process as well as the parameters of the BTR strategy. In our model, we start with the distribution of the number of arrivals during a vacation time based on an event tree, and then derive the power efficiency and a generalized P-K formula of the mean delay for EEE systems.

Our analysis shows that the counter and the timer play different roles in performance guarantees and compensate each other to adapt to traffic fluctuations. Therefore, the $\tau\&N$ policy can better adapt to the traffic on Ethernet networks, which typically are bursty. Based on these properties, we provide the rules to select appropriate values for parameters $\tau$ and $N$. The methodology developed in this paper can be generalized and applied to the analysis of queueing models with vacation time that is governed by the arrival process.

\appendices
\section{Failure of the Classical P-K Formula in the vacation model of EEE Protocol}

\begin{figure} [b]
\centering
\includegraphics[scale=0.7]{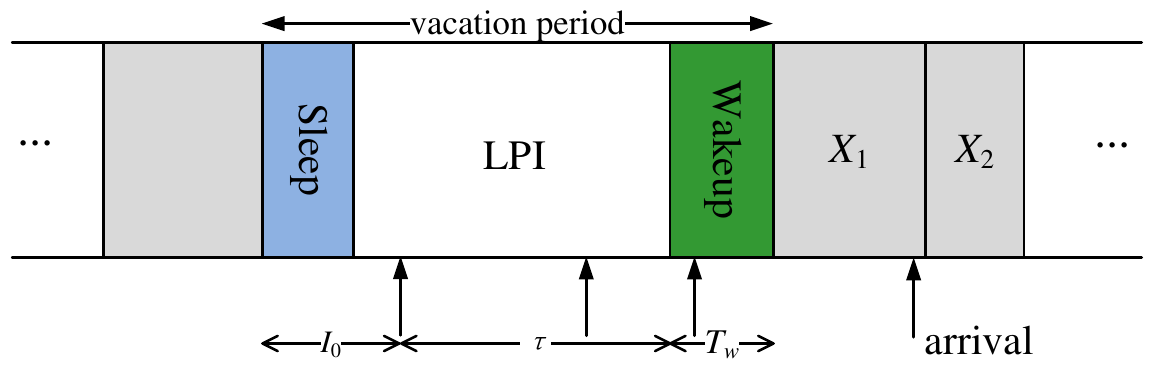}
\caption{Illustration of the $\tau$ policy.}
\label{AppTpolicy}
\end{figure}

It is well-known that, with a Poisson arrival process, the relation $$H(z)=V^*(\lambda-\lambda z)$$ holds in the classical M/G/1 queue with vacation time,where $H(z)$ is the $z$ transform of the distribution $h_n$ of the number of the arrivals during the vacation and $V^*(s)$ is the Laplace transform of the vacation time distribution $v(x)$. However, this relation is invalid in the vacation model of the EEE protocol, where the vacation time is regulated by the arrival process. In the following, we take the $\tau$ policy as an example to demonstrate this point.

As Fig. \ref{AppTpolicy} shows, the vacation time of the $\tau$ policy consists of three parts: the time from the beginning of the vacation to the first arrival $I_0$, the LPI time $\tau$, and the Wakeup period $T_w$. As $\tau$ and $T_w$ are constants, the probability distribution function (PDF) of vacation time is given by:
\begin{equation}
v(x)=\lambda e^{-\lambda(x-\tau-T_w)},\  x\geq\tau+T_w.
\end{equation}
Taking the Laplace transform, we have:
\begin{align}
V^*(s)&=\int_0^\infty e^{-sx}v(x)dx \nonumber \\
&=\lambda \int_{\tau+T_w}^\infty e^{-sx}e^{-\lambda(x-\tau-T_w)}dx \nonumber \\
&=\frac{\lambda}{\lambda+s}e^{-s(\tau+T_w)}.
\end{align}
On the other hand, the distribution of the number of arrivals during vacation time $h_n$ is given by:
\begin{equation}
h_n= e^{-\lambda(\tau+T_w)}\frac{[\lambda(\tau+T_w)]^{n-1}}{(n-1)!},\ (n\geq1).\,
\end{equation}
and the generating function of $h_n$ can be expressed as follows:
\begin{equation}
H(z)=\sum_{n=0}^\infty h_nz^n=ze^{-\lambda(1-z)(\tau+T_w)}.
\end{equation}
Clearly, the equality $H(z)=V^*(\lambda-\lambda z)$ does not hold in the vacation model of the EEE protocol.

Next, we examine whether the P-K formula (19) for the traditional M/G/1 queue with vacation time can still be applied to the vacation model of the EEE protocol. From (43), the first and the second moment of the vacation time are respectively given as follows:
\begin{equation}
\overline{V}=-{V^{*}}'(0)=\frac{1}{\lambda}+\tau+T_w,
\end{equation}
\begin{equation}
\overline{V^2}={V^*}''(0)=\frac{2}{\lambda^2}+(\tau+T_w)^2.
\end{equation}
Substituting (46) and (47) into the classical P-K formula (19), we obtain:
\begin{align}
D&=\frac{\lambda\overline{X^{2}}}{2(1-\rho)}+\frac{\overline{V^{2}}}{2\overline{V}}+\overline{X}\nonumber \\
&=\frac{\lambda\overline{X^{2}}}{2(1-\rho)}+
\frac{\frac{2}{\lambda^2}+\frac{2(\tau+T_w)}{\lambda}+(\tau+T_w)^2}{2(\frac{1}{\lambda}+\tau+T_w)}
+\overline{X}.
\end{align}

\begin{figure}
\centering
\includegraphics[scale=1.25]{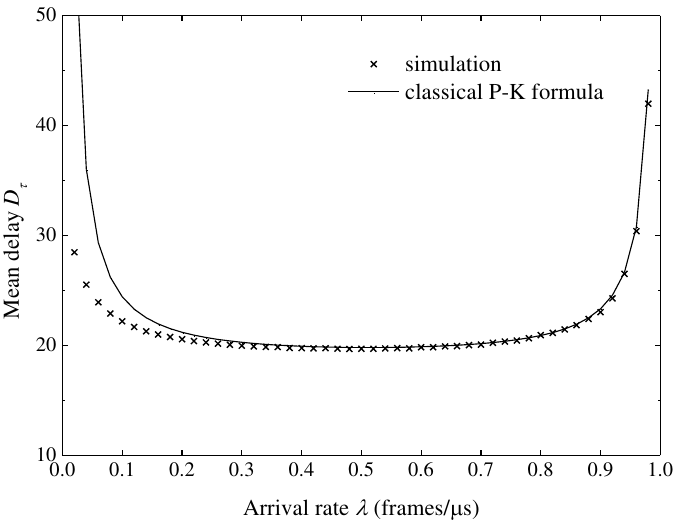}
\caption{Validation of the classical P-K formula, where $\tau=30 \mu s$.}
\label{validation}
\end{figure}

We verify the above result (48) by simulation. Fig. \ref{validation} shows that the simulation results are always lower than the numerical results of equation (48). This discrepancy reveals the difference between the classical M/G/1 queue with vacation time and the vacation model of the EEE protocol. In the former system, the vacation time is independent of the arrival process. For a packet that arrived in a vacation period, the density of the residual vacation time is given by \cite{kleinrock1975theory}:
\begin{equation}
\hat{f_v}(x)=\frac{1-F_v(x)}{\overline{V}},
\end{equation}
where $F_v(x)$ is the distribution of the vacation time $V$. However, in an EEE protocol with the $\tau$ policy, the residual vacation time of the first arrival is deterministic and equals $\tau+T_w$. Therefore, the density of the residual vacation time in the model of EEE protocol is no longer given by (49). As Fig. \ref{validation} shows, the discrepancy is more significant when the arrival rate $\lambda$ is smaller because fewer packets arrive in a vacation time.

\ifCLASSOPTIONcaptionsoff
  \newpage
\fi

\bibliographystyle{ieeetr}
\bibliography{IEEEabrv,refPXD}
\end{document}